\documentstyle{article}
\title{Quantum field theory on manifolds with a boundary}
\author{ Z. Haba\\Institute of Theoretical Physics, University of Wroclaw,
\\50-204 Wroclaw, Plac Maxa Borna 9, Poland\\e-mail:zhab@ift.uni.wroc.pl}
\date{PACS numbers 04.62+v,02.50Cw;Key words:random fields, boundary quantum
fields}
\begin{document}
\maketitle
\begin{abstract}
We discuss quantum  theory of fields $\phi$ defined on a
$(d+1)$-dimensional manifold ${\cal M}$ with a boundary ${\cal
B}$. The free action $W_{0}(\phi)$ which is  a bilinear form in
$\phi$ defines the Gaussian measure  with a covariance (Green
function) ${\cal G}$ . We discuss a relation between the quantum
field theory with a fixed boundary condition $\Phi$ and the theory
defined by the Green function ${\cal G}$. It is shown that the
latter results by an average over $\Phi$ of the first. The QFT in
AntiDeSitter space is treated as an example. It is shown that
quantum fields on the boundary are more regular than the ones on
AntiDeSitter space.
\end{abstract}
   \section{Introduction}
    Quantum field theory (QFT) can be  defined by a
    functional integral
    \begin{equation}
    d\mu(\phi)= {\cal D}\phi \exp(-W(\phi))
    \end{equation}
    where $W$ is the classical action. From the point of view of
    formal properties (translational invariance) of such a functional integral
     it should not
    matter whether we write in it $\phi$ or $\phi+\phi_{0}$.
 However, if the action is defined on a manifold with a boundary
then the dependence on the boundary value of $\phi $ seems to be
crucial [1]-[3]. This means that a formal invariance under
translations in function space $\phi\rightarrow \phi+\phi_{0}$
must be broken in the definition of the functional integral in
refs.[1]-[3]. Then, the dependence on the boundary value breaks
some symmetries present in the classical action $W$.
 Such an approach to QFT disagrees with the conventional one
based  on the mode summation \cite{ford}\cite{fronsdal}
\cite{bunch}\cite{schlom}\cite{davis}\cite{guth}\cite{tsamis}\cite{ryang}
or perturbation expansion in the number $N$ of components or in
the coupling constant.
 In this paper we discuss a relation between the two approaches in
 the framework of the functional integral. In the AntiDeSitter
 models of refs.[1]-[3] the boundary appears at the spatial infinity and
  coincides with the (compactified)
 Minkowski space. In the Euclidean version of the AntiDeSitter
 space (in the Poincare coordinates) the boundary can be realized
as the Euclidean subspace of the hyperbolic  space.

 We consider a Riemannian manifold ${\cal M}$ with the
boundary ${\cal B}$. The metric on ${\cal M}$ is denoted by $G$
and its restriction to ${\cal B}$ by $g$. We shall denote the
coordinates on ${\cal M}$ by $X$ and their restriction to ${\cal
B}$ by $x$; close to the boundary we write $X=(y,x)$. The action
for a minimally coupled massless free scalar field $\phi$ reads
\begin{equation}
W_{0}(\phi)=\int_{{\cal M}}
dX\sqrt{G}G^{AB}\partial_{A}\phi\partial_{B}\phi \equiv
(\phi,{\cal A}\phi)\end{equation} The non-negative bilinear form
(2) is defined on a certain domain $D({\cal A}) $ of functions.
Such a bilinear form determines a self-adjoint operator ${\cal A}$
\cite{kato}(the definition of ${\cal A}$ depends on the choice of
$D({\cal A})$) in the Hilbert space of square integrable
functions. The free (Euclidean) quantum field $\phi$ is defined by
${\cal A}^{-1}$ in the sense that the kernel of ${\cal A}^{-1}$
(the Green function) provides a definition of the two-point
correlation function of $\phi$. The Green function ${\cal G}$ is a
solution of the equation
\begin{equation}
 -{\cal A}{\cal G}\equiv\partial_{A}G^{AB}\sqrt{G}\partial_{B} {\cal G}=\delta
\end{equation}
where $G=\det{G_{AB}}$. The solution of eq.(3) is not unique . If
${\cal G}^{\prime}$ is another solution of eq.(3) then ${\cal
G}^{\prime}={\cal G}+{\cal S}$ where ${\cal S}$ is a solution of
the equation
\begin{equation}
{\cal A}{\cal S}=0
\end{equation}
We can determine ${\cal G}$ unambiguously imposing some additional
requirements, e.g., requiring that ${\cal G}=0$ on the boundary.
The various definitions of ${\cal G}$ correspond to various
choices of $D({\cal A})$ in the definition of the bilinear form
(2)\cite{kato}\cite{guerra}.

\section{The functional measure}

 To the free action (2) we add a local interaction $V$. Now, the
 total action reads
\begin{equation} W=W_{0}+W_{I}=\int_{{\cal M}}
dX\sqrt{G}G^{AB}\partial_{A}\phi\partial_{B}\phi +\int_{{\cal M}}
dX\sqrt{G} V(\phi)
\end{equation}
We can give a mathematical definition of the formal  functional
measure (1)
\begin{equation}
d\mu_{V}(\phi)= Z_{0}^{-1}d\mu_{0}(\phi)\exp(-W_{I})
\end{equation}
where the Gaussian measure $\mu_{0}$ is a mathematical realization
of the formal integral
\begin{displaymath} d\mu_{0}(\phi) ={\cal
D}\phi\exp(-W_{0})\end{displaymath} The partition function $Z_{0}$
in eq.(6)
\begin{equation} Z_{0}=\int
d\mu_{0}\exp(-W_{I})
\end{equation} determines a normalization factor.

We do not discuss in this paper some divergence problems which may
arise if $V$ is a local function of $\phi$ and ${\cal M}$ has an
infinite volume. We may assume that $W_{I}$ has been properly
regularized. We have a suggestion how to construct a regular QFT
at the end of this paper.

A functional measure $\mu$ defines a probability distribution of
fields $\phi(X)$; more precisely the "smeared out " fields
\begin{displaymath}
(\phi,f)=\int dX\sqrt{G}\phi(X)f(X).\end{displaymath} In
probability theory (see, e.g., \cite{simon}\cite{cramer}) it is
convenient to treat the probability measure $\mu$ defined on some
sets of random fields $\phi$ as one of many possible realizations
of the probability space $(\Omega,\Sigma,P)$. The random field
$\phi:\Omega\rightarrow R$ (at fixed $X$) is a map from the set
$\Omega$ to the set of real numbers such that the two-point
correlation function is an average over the "sample paths"
$\omega\in\Omega$
\begin{displaymath}
\langle \phi(X)\phi(Y)\rangle=\int_{\Omega}
dP(\omega)\phi_{\omega}(X)\phi_{\omega}(Y)
\end{displaymath}
where $P$ is a probability measure on the $\sigma$-algebra
$\Sigma$ of subsets of $\Omega$. The Gaussian measure gives a
realization of the Gaussian random field $\phi_{\omega}$. It  is
defined \cite{simon}\cite{skorohod} by the mean
\begin{displaymath}
m(X)=\int d\mu(\phi)\phi(X)\equiv \langle \phi(X)\rangle
\end{displaymath} and the covariance
\begin{equation}
\begin{array}{l}
{\cal G}(X,Y)=\int d\mu(\phi)(\phi(X)-\langle
\phi(X)\rangle)(\phi(Y)-\langle \phi(Y)\rangle)\cr
=\langle(\phi(X)-\langle \phi(X)\rangle)(\phi(Y)-\langle
\phi(Y)\rangle)\rangle
\end{array}\end{equation}
or by its characteristic function $S$
\begin{displaymath}
S[if]=\int d\mu\exp(i(\phi,f))=\exp(i(m,f)-\frac{1}{2}(f,{\cal
G}f))
\end{displaymath}
 Note that if we make a shift in
the function space and define $\tilde{\phi}=\phi-m$ then
$\tilde{\phi}$ has zero mean. Hence, we could subtract the mean
value defining a new Gaussian measure
\begin{displaymath}
d\tilde{\mu}(\tilde{\phi})=d\mu(\tilde{\phi}+m)
\end{displaymath}
The Gaussian measure is quasiinvariant under a shift $\chi$ if
there exists an integrable function $\rho(\phi,\chi)$ such that
\begin{equation} d\mu(\phi
+\chi)=d\mu(\phi)\rho(\phi,\chi)
 \end{equation}
 It is easy to see by a calculation of the characteristic function of both sides
 of eq.(9) that the measure $\mu$ is quasiinvariant under the
 shift $\chi$ if \cite{skorohod}
 \begin{equation}
 \rho(\phi,\chi)=\exp(-(\phi,B\chi)-\frac{1}{2}(\chi,C\chi))
 \end{equation}
 and the following equations are satisfied
 \begin{equation}
 \chi={\cal G}B\chi
 \end{equation}
 \begin{equation}
 (B\chi,{\cal G}B\chi)=(\chi,C\chi)
 \end{equation}
Eqs.(9)-(12) express the formal invariance  of the functional
measure (1) under translations in the function space. If these
conditions are not satisfied then it really does matter what is
the shift $\chi$. In some papers on AdS-CFT correspondence
[1]-[3],\cite{arefeva}\cite{ryang} the choice is made ${\cal
G}(X,Y)={\cal G}_{D}(X,Y)$ where ${\cal G}_{D}$ is the Dirichlet
Green function (vanishing on the boundary) and $\langle
\phi(X)\rangle=\phi_{0}(X)$ where $\phi_{0}(X)$ is a solution of
 the  equation
\begin{equation}
{\cal A}\phi_{0}=0
\end{equation}
 with a fixed boundary condition $\Phi$. We can see that eqs.(11)-(12)
 cannot be satisfied
 if $\chi=\phi_{0}$.
 Hence, the partition function $Z[\Phi] $ may depend on the boundary value $\Phi$.

 In general,
choosing in QFT the boundary field $\phi_{0}\neq 0$  we break some
symmetries of the classical action (5). As an example
 we could consider the hyperbolic space with the metric
\begin{equation}
 ds^{2}=y^{-2}(dy^{2}+ dx_{1}^{2}
 +....+dx_{d}^{2})
 \end{equation}
 The hyperbolic space (14) has compactified $R^{d}$ as the boundary \cite{witten}.
 The hyperbolic space can be considered as an Euclidean
version of AntiDeSitter space ($AdS_{d+1}$ has compactified
Minkowski space as a boundary at conformal infinity
\cite{witten}). It is also a Euclidean version of DeSitter space.
However, the Poincare coordinates (14) are inappropriate for an
analytic continuation of quantum fields from the hyperbolic space
to DeSitter space (there is also no boundary at conformal infinity
of DeSitter space).

 The
action (5) in the hyperbolic space is invariant under $R^{d}$
rotations and translations whereas the quantum field theory with a
fixed boundary value of $\phi_{0}$ would not be invariant under
these symmetries.
 The approach to QFT assuming  a boundary condition $\Phi$ for the field
 $\phi_{0}$ and the Dirichlet
boundary condition for the Green function leads to a different
quantum field theory than the one developed in
refs.\cite{fronsdal}\cite{bunch}\cite{schlom}\cite{davis}. The
latter
 is determined by the mean $\langle \phi\rangle =0$
 and a choice of the Green function (the free propagator
 ${\cal G}$
solving eq.(3)) which does not vanish on the boundary. A possible
way to determine the propagator is to construct it for a real time
by a mode summation and subsequently to continue analytically the
propagator to the imaginary time (for a class of models  this is
done in \cite{ford}\cite{bunch}; the mode summation is also not
unique). It seems reasonable
 to choose the Green function ${\cal G}$ which has the symmetries
 of the action $W_{0}$ (1)as in \cite{ford}\cite{fronsdal}\cite{bunch}\cite{schlom}\cite{ryang}. Then, the functional measure (6)
 will have the symmetries of the action (5).

\section{ An average over the boundary values}
After the heuristic discussion in sec.2 of functional integration
over fields with a fixed boundary value
 we prove in this section that the approach starting form the free propagator
 ${\cal G}$ is equivalent to a quantization
 around a classical  solution $\phi_{0}$ with a prescribed boundary value $\Phi$
 if subsequently an average over
 all such boundary values is performed.
First, let us assume (in the sense that for the bilinear forms
$(f,{\cal G}f)\geq (f,{\cal G}_{D}f)$)
\begin{equation}
{\cal G}\geq {\cal G}_{D}
\end{equation}
If the operator $ {\cal A}$ is an elliptic operator then the
inequality (15) follows  from the maximum principle for elliptic
operators \cite{jaffe}\cite{gilbarg}. We are interested also in
operators ${\cal A}$ with singular or vanishing coefficients which
need not be elliptic. It is not clear whether the inequality (15)
can be satisfied for such operators. However, the inequality (15)
still holds true for the Green functions of  singular operators
discussed in \cite{haba1}\cite{haba2} which are expressed by a
path integral. The Dirichlet condition imposes a restriction on
the class of paths. Hence, the integral over a restricted set of
paths is bounded by ${\cal G}$ in eq.(15).

 If the inequality (15)
is satisfied then there exists a positive definite bilinear form
${\cal G}_{B}$ such that
 \begin{equation} {\cal G}(X,X^{\prime})=
G_{D}(X,X^{\prime})+{\cal G}_{B}(X,X^{\prime})
\end{equation}
Clearly on the boundary
\begin{equation}{\cal
G}(0,x;0,x^{\prime})={\cal G}_{B}(0,x;0,x^{\prime})\equiv{\cal
G}_{E}(x,x^{\prime})
\end{equation}
${\cal G}_{E}$ defines a non-negative  bilinear form on the set of
functions defined on the boundary ${\cal B}$.

{\bf Theorem 1}

Let $\mu_{0}$ be the Gaussian measure with the mean zero and the
covariance ${\cal G}$. Assume that  ${\cal G}$ and ${\cal G}_{D}$
are real positive definite bilinear forms satisfying the
inequality (15). Then, there exist independent Gaussian random
fields $ \phi_{D}$ and $\phi_{B}$ with the mean equal zero and the
covariance  ${\cal G}_{D}$ and ${\cal G}_{B}$ resp. such that for
any integrable function $\exp(-W_{I})F$
 \begin{equation} \begin{array}{l}\int
d\mu_{0}(\phi)\exp(-W_{I}(\phi))F(\phi)\cr =\int
d\mu_{D}(\phi_{D})
d\mu_{B}(\phi_{B})\exp(-W_{I}(\phi_{D}+\phi_{B}))F(\phi_{D}+\phi_{B})
\end{array}\end{equation}
In this sense
\begin{equation}
\phi=\phi_{D}+\phi_{B}
\end{equation}

The theorem and its proof can be found in
\cite{guerra}\cite{simon}. It is easy to check eq.(18) for the
generating functional ( then $\exp(-W_{I}(\phi))F(\phi)=\exp(
\phi,J)$). On a perturbative level the general formula (18)
follows from the one for the generating functional. For the
general theory of "conditioning" (15) see
\cite{guerra}\cite{simon}. Eq.(18) is discussed in the lattice
approximation in \cite{simon} (sec.8.1). Another derivation and a
discussion of its relevance to the AdS-CFT correspondence can be
found in \cite{rehren2}. The relevance of an average over the
boundary values for the Hamiltonian formulation of the quantum
field theory is discussed in \cite{marlow}.

Let us note that on a formal level
\begin{displaymath}
d\mu_{D}(\phi_{D})={\cal
D}\phi_{D}\exp(-\frac{1}{2}(\phi_{D},{\cal A}_{D}\phi_{D}))
\end{displaymath}
where ${\cal A}_{D}$ is the Laplace-Beltrami operator with the
Dirichlet boundary conditions. On a formal level  ${\cal
A}_{D}\phi_{0}=0$. Hence, $d\mu_{D}(\phi_{D}+\phi_{0})=
d\mu_{D}(\phi_{D}) $ although strictly speaking the shift of
$\mu_{D}$ by $\phi_{0}$ does not make sense because $\phi_{0}$
does not vanish on the boundary. We treat the r.h.s. of eq.(18)
(before an integration over $\phi_{B}$) as a rigorous version
 of the  QFT  shifted by a classical solution.
 This interpretation is suggested by

{\bf Theorem 2}

 Let  ${\cal G}_{D}$ be  the Dirichlet Green
function of the operator ${\cal A}$ (eq.(3)). Let ${\cal G}$ be
another real solution of eq.(3) satisfying the inequality (15).
Then, there exists a Gaussian random field $\Phi$ defined on the
boundary ${\cal B}$ with the mean zero and the  covariance ${\cal
G}_{E}$ such that ( in the sense of $L^{2}(dP)$ integrals
\cite{cramer})
\begin{equation}
\phi_{B}(X)=\int_{{\cal B}} dx_{b}\sqrt{g}{\cal
D}(X,x_{b})\Phi(x_{b})
\end{equation}
where ${\cal D}(X,x_{b})$ is the Green function solving the
boundary value problem for eq.(13).

 {\bf Proof}: Let us note that
${\cal G}_{D}$ as well as ${\cal G}$ satisfy the same equation
(3). Then, their difference ${\cal G}_{B}={\cal G}-{\cal G}_{D}$
satisfies the equations
\begin{equation}
{\cal A}(X){\cal G}_{B}(X,X^{\prime})={\cal A}(X^{\prime}){\cal
G}_{B}(X,X^{\prime})=0
\end{equation}
and the boundary condition ${\cal G}_{B}(0,x;0,x^{\prime})={\cal
G}_{E}(x,x^{\prime})$. We can solve eq.(21) with the given
boundary condition ${\cal G}_{E}$
\begin{equation}
{\cal G}_{B}(X,X^{\prime})=\int_{{\cal B}} dx_{b}\sqrt{g}{\cal
D}(X,x_{b})\int_{{\cal B}} dx_{b}^{\prime}\sqrt{g} {\cal
D}(X^{\prime},x_{b}^{\prime}){\cal G}_{E}(x_{b},x_{b}^{\prime})
\end{equation}
where ${\cal D}$ is the Green function solving the Dirichlet
boundary  problem for eq.(13). The bilinear form ${\cal G}_{E}$
(17) defines a Gaussian field $\Phi$ on the probability space
$(\Omega,\Sigma,P)$ (see sec.2) with the mean zero and the
covariance
\begin{equation}
\langle \Phi(x)\Phi(x^{\prime})\rangle ={\cal G}_{E}(x,x^{\prime})
\end{equation}
We define $\tilde{\phi}_{B}$ by the r.h.s. of eq.(20) where the
integral can be understood in the sense of the $L^{2}(dP)$
convergence of the Riemann sums (see, e.g.,\cite{cramer}). For the
proof of the theorem ($\tilde{\phi}=\phi$) it is sufficient to
show that the covariance of $\tilde{\phi}_{B}$ coincides (as a
bilinear form) with ${\cal G}_{B}$. This is a consequence of
eq.(22).

Let us note that there exists the Gaussian measure $\nu_{B}$ such
that \begin{displaymath}
 \int d\nu_{B}(\Phi)
\Phi(x)\Phi(x^{\prime})={\cal G}_{E}(x,x^{\prime})
\end{displaymath}
$\nu_{B}$ can be defined by $\mu_{B}$ as $\nu_{ B}=\mu_{ B}\circ
T$ where $T(\Phi)=\phi_{B}$ is the one to one map (20) expressing
the solution of eq.(13) by its boundary value. We can see that if
there is a QFT with a two-point function ${\cal G}$ non-vanishing
on the boundary then there is the unique choice of $\phi_{0}$
solving eq.(13) such that $\phi=\phi_{D}+\phi_{0}$ is a
realization of a random field with the boundary value $\Phi$. An
average over $\Phi$ leads to the Green function ${\cal G}$.

In the example of  the hyperbolic space (14) (with the Poincare
coordinates) the solution (20)
 of the Dirichlet boundary problem (13) can be expressed
 by its boundary value $\phi_{B}(y=0,x)=\Phi(x)$\cite{arefeva}
 \begin{equation}
 \phi_{B}(X)=y^{\frac{d}{2}}\int dp\exp(ipx)
 \vert p\vert^{\frac{d}{2}}K_{\frac{d}{2}}(\vert p\vert
 y)\tilde{\Phi}(p)\end{equation}
 where $\tilde{\Phi}$ denotes the Fourier transform of $\Phi$
 and  $K_{\nu}$ is the modified Bessel function of order $\nu$ \cite{stegun}.
Comparing with eq.(20) we obtain ($X=(y,x)$)
\begin{displaymath}
{\cal D}(X,x^{\prime})=(2\pi)^{-d}y^{\frac{3}{2}d+1}\int
dp\exp(ip(x-x^{\prime}))\vert
p\vert^{\frac{d}{2}}K_{\frac{d}{2}}(\vert p\vert
 y)
\end{displaymath}
 The two-point function $ {\cal G}_{E}$ resulting from the QFT on the hyperbolic space constructed
 in refs.\cite{ford}\cite{bunch}\cite{tsamis} is ${\cal G}_{E}=-\ln \vert x-x^{\prime}\vert$
 \cite{guth}\cite{hawking}\cite{allen}\cite{dolgov}.  In the Fourier transforms
(up to an inessential normalization) we have ( see the discussion
in
\cite{ford}\cite{dolgov}\cite{haba1}\cite{haba2})\begin{equation}
 \langle \tilde{\Phi}(p)\tilde{\Phi}^{*}(p^{\prime})\rangle=
 \delta(p-p^{\prime})\vert p\vert^{-d}
 \end{equation}
We may apply eq.(25) to calculate $\langle
\phi_{B}(X)\phi_{B}(X^{\prime})\rangle $. As a solution of eq.(21)
after an analytic continuation to the real time it must coincide
with the Hadamard two-point function (vacuum expectation value
 of an anticommutator of quantum scalar fields)
 which is usually denoted by $G^{(1)}(X,X^{\prime})$ ( the formula
for ${\cal G}_{B}$ in the hyperbolic space can be found in
\cite{allen} and for ${\cal G}_{D}$ in \cite{maldacena2} )
\begin{equation}
{\cal G}_{B}(X,X^{\prime})= \langle
\phi_{B}(X)\phi_{B}(X^{\prime})\rangle=(yy^{\prime})^{\frac{d}{2}}\int
dp \exp(ip(x-x^{\prime}))K_{\frac{d}{2}}(\vert p\vert y
)K_{\frac{d}{2}}(\vert p\vert y^{\prime} )
\end{equation}
\section{Non-linear boundary value problem}
 We can modify the formulation (6)-(13) of QFT on
manifolds with a boundary so that  the interaction $V(\phi)$ is
taken into account already at the classical level. Then, instead
of eq.(13) we consider the equation
\begin{equation}
-{\cal A}\psi=V^{\prime}(\psi)
\end{equation}
or in the integral form
\begin{equation} \psi(X)=\phi_{B}(X)+ \int dX^{\prime}\sqrt{G}
{\cal G}_{D}(X,X^{\prime})V^{\prime}(\psi(X^{\prime}))
\end{equation}
where $\phi_{B}$ is defined in eq.(20). In order to express the
functional integral (6) in terms of $\psi$ let us introduce a
differential operator
\begin{equation}
{\cal A}_{\psi}= {\cal A }+ V^{\prime\prime}(\psi)
\end{equation}
Define ${\cal G}_{D}^{\psi}$ as the Dirichlet Green function of
${\cal A}_{\psi}$. Let $\mu_{\psi}$ be the Gaussian measure with
the mean zero and the covariance ${\cal G}_{D}^{\psi}$. Then, the
formula (18) reads (under the assumption that the function on the
r.h.s. of eq.(30) is integrable)
\begin{equation}
\begin{array}{l}
\int d\mu_{0}(\phi)\exp(-W_{I}(\phi))F(\phi)=\int
d\mu_{B}(\phi_{B})\exp(W_{0}(\phi_{B})-W(\psi))\det({\cal
A}_{\psi})^{-\frac{1}{2}}\det({\cal A})^{\frac{1}{2}}\cr\int
d\mu_{\psi}(\phi_{D})
  \exp\Big(-\int
dX\sqrt{G}V(\phi_{D}+\psi)+\int dX\sqrt{G}V(\psi)+\int dX\sqrt{G}
V^{\prime}(\psi)\phi_{D} \cr +\frac{1}{2}\int dX\sqrt{G}
\phi_{D}V^{\prime\prime}(\psi)\phi_{D}\Big)F(\phi_{D}+\psi)
\end{array}
\end{equation}
For the proof let us shift variables in eq.(18) and apply
eqs.(9)-(12). Then,
\begin{equation}
\begin{array}{l}
d\mu_{D}(\phi_{D}+\chi)\exp(-\int dX
\sqrt{G}V(\phi_{D}+\phi_{B}+\chi)) F(\phi_{D}+\phi_{B}+\chi)\cr
=d\mu_{D}(\phi_{D}))\exp(-\frac{1}{2}\int dX\sqrt{G}\chi{\cal
A}\chi)F(\phi_{D}+\psi) \cr \exp\Big(-\int dX\sqrt{G}\chi {\cal
A}\phi_{D} -\int dX\sqrt{G}V(\phi_{D}+\psi)\Big)
\end{array}
\end{equation}
where in the second step we inserted $\chi=\psi-\phi_{B}$
($\chi=0$ on the boundary, hence the shift is admissible). Next,
we make use of ${\cal A}\phi_{B}=0$ (then ${\cal A}\psi={\cal
A}\chi$), subtract the two first terms of the Taylor expansion of
$V(\phi_{D}+\psi)$ in $\psi$ and apply the formula for a Gaussian
integral of an exponential of a quadratic form \cite{skorohod}

\begin{equation}
d\mu_{D}(\phi_{D}))\exp\Big(-\frac{1}{2}\int
dX\sqrt{G}\phi_{D}V^{\prime\prime}(\psi)\phi_{D}\Big)= (\det {\cal
A}_{\psi})^{-\frac{1}{2}} d\mu_{\psi}(\phi_{D})
\end{equation}
 The final result is expressed in eq.(30). In this equation
 $\exp(-W(\psi)) $ is the effective action in the tree
 approximation
 (discussed by \cite{ryang}) and $\det {\cal
 A}_{\psi}^{-\frac{1}{2}}$ gives the one-loop approximation
 to the effective action in QFT with the boundary value $\Phi$ . The remaining
 $d\mu_{\psi}(\phi_{D})$ integral in eq.(30) can be calculated in perturbation expansion.
 It starts with higher powers $n$ ($n\geq 3$) of $\phi_{D}$
 leading to  corrections in higher loops to the effective action.

\section{Conclusions}
In this section we  derive some relations between correlation
functions with respect to various measures discussed in earlier
sections. Let us define
\begin{equation}
Z[\Phi]=\exp(-W_{0}(\phi_{B}))\int
d\mu_{D}(\phi_{D})\exp(-W_{I}(\phi_{D}+\phi_{B}))
\end{equation}
where $\phi_{B}$ is defined in eq.(20) with $\Phi$ as a fixed
boundary value. The definition (33) is introduced in such a way
that it  agrees with the large $N$ formula of \cite{polyakov} and
the semiclassical calculations of \cite{ryang} and the ones in
eq.(30) (see also a discussion in  \cite{rehren1}).

 If $Z[\Phi]$ is the generating
functional then there exists a field ${\cal O}({\bf x})$ such that
\begin{equation}
 Z[\Phi]=\langle\exp(\int_{{\cal B}}{\cal
 O}(x)\Phi(x)\sqrt{g}dx)\rangle
 \end{equation}
 Treating $Z[\Phi]$ as the generating functional we can calculate
\begin{equation}
\begin{array}{l}
\frac{\delta}{\delta \Phi({\bf x}_{1})}....\frac{\delta}{\delta
\Phi({\bf x}_{n})}Z[\Phi]_{\vert \Phi=0}  =( {\cal
D}\frac{\delta}{\delta \phi_{B}})({\bf x}_{1})....({\cal
D}\frac{\delta}{\delta \phi_{B}})({\bf x}_{n})\cr
\exp(-W_{0}(\phi_{B}))\int d\mu_{D}(\phi_{D})
\exp(-W_{I}(\phi_{D}+\phi_{B}))_{\vert \phi_{B}=0}
\end{array}
\end{equation}
where \begin{displaymath} ({\cal D}\frac{\delta}{\delta
\phi_{B}})({\bf x})\equiv \int dX\sqrt{G}{\cal D}(X,{\bf
x})\frac{\delta}{\delta \phi_{B}(X)} \end{displaymath} and
\begin{displaymath}
W_{0}(\phi_{B})=\frac{1}{2}(\phi_{B},{\cal A}\phi_{B})
\end{displaymath}
We wish to compare these correlation functions with the ones of
the bulk field $\phi$ defined by the generating functional
\begin{equation}
S[J]=\int d\mu_{0}(\phi)\exp(-W_{I}(\phi)+(J,\phi))
\end{equation}
Then, the correlation functions can be calculated from the formula
\begin{equation}
\begin{array}{l}
\frac{\delta}{\delta J(X_{1})}....\frac{\delta}{\delta
J(X_{n})}Z[J]_{\vert J=0}=({\cal G} \frac{\delta}{\delta
\phi_{c}})(X_{1})....({\cal G}\frac{\delta}{\delta
\phi_{c}})(X_{n})\cr\exp(\frac{1}{2}(\phi_{c},{\cal
A}\phi_{c}))\int d\mu_{0}(\phi) \exp(-W_{I}(\phi+\phi_{c}))_{\vert
\phi_{c}=0}
\end{array}\end{equation}
where on the r.h.s. we have absorbed the linear term of the
exponential (36) into a shift of the measure according to
eqs.(9)-(10) with $\phi_{c}={\cal G}J$ and $J={\cal A}\phi_{c}$ .
It is clear from eqs.(35) and (37) that a perturbative calculation
of $n$-point  correlation functions of ${\cal O}$ and $\phi$
involves the same graphs and only the propagators are different.
 The relation between (35) and (37) has been discovered earlier
by Duetsch and Rehren \cite{rehren2} ( see also  \cite{banks}). A
Hamil

tonian derivation of the relation between differentiation with
respect to boundary values and sources $J$ can be found in
\cite{marlow}.

 The field theory in the bulk (6)-(7) is an integral over $Z[\Phi]$
 \begin{equation}
 Z_{0}=\int d\nu_{B}(\Phi)\exp(W_{0}(\phi_{B})) Z[\Phi]
 \end{equation}
 We can obtain a connection between some
other correlation functions. Generalizing eq.(38) let us define
the generating functional $S_{D}[\phi_{B};J]$ in the $\phi_{D}$
theory shifted by a background field $\phi_{B}$
 \begin{equation}
 S_{D}[\phi_{B};J]=\int d\mu_{D}(\phi_{D})
 \exp(-W_{I}(\phi_{D}+\phi_{B}))\exp(\int dX\sqrt{G}J\phi_{D})
\end{equation}
Then, from eq.(18) the generating functional for correlation
functions of the fields $\phi$ in the model (6) is
\begin{equation}
S[J]=\int d\mu_{B}(\phi_{B})\exp(\int dX\sqrt{G}J\phi_{B}) S_{D}[
\phi_{B};J]
\end{equation}
It can be seen that $\phi_{D}$ and $\phi_{B}$ enter symmetrically
in eq.(18). Hence, we may also write
\begin{equation}
S[J]=\int d\mu_{D}(\phi_{D})\exp(\int dX\sqrt{G}J\phi_{D}) S_{B}[
\phi_{D};J]
\end{equation}
Differentiating  both sides of eq.(40) and (41) we obtain a
relation between correlation functions of the fields
$\phi$,$\phi_{D}$ and $\phi_{B}$. The form of the correlation
functions in the model (6) at the boundary points ${\bf x}_{j}\in
{\cal B}$ is a simple consequence of eq.(41)
\begin{equation}
\langle \phi({\bf x}_{1})...\phi({\bf x}_{n})\rangle =\int
d\mu_{D}(\phi_{D})\int d\nu_{B}(\Phi)
\exp(-W_{I}(\phi_{D}+\phi_{B}))\Phi({\bf x}_{1})...\Phi({\bf
x}_{n})
\end{equation}
In particular, if the interaction is concentrated only on the
boundary
\begin{displaymath}
W_{I}(\phi)=\int_{{\cal B}} d{\bf x}\sqrt{g}V(\phi(0,{\bf x}))
\end{displaymath}
then $\phi_{D}=0$ in $W_{I}$ in eq.(42) and the functional
integral (42) is the same as in QFT on ${\cal B}$ defined by the
free field measure $d\nu_{B}$ with the covariance ${\cal
G}_{E}({\bf x},{\bf x}^{\prime})$. In the case of the hyperbolic
space this covariance is logarithmic. Hence, ultraviolet problem
is the same as for quantum fields in two dimensions.

 We think that the  QFT theory on
a boundary of a curved manifold is interesting for itself because
of its remarkable regularity expressed  (for the hyperbolic space)
in the strong decay (25) in the momentum space. However, the main
result of this paper is formulated in eqs.(40)-(42 ). The formulas
connecting the correlation functions of fields in various field
theoretic models can shed some light on relations of the AdS-CFT
type.

{\bf Acknowledgements}

The author is grateful to an anonymous referee for pointing out
ref.[22]. The formulas at the beginning of sec.5 resulted from my
effort to derive the results of Duetsch and Rehren in the
continuum as the addendum to an earlier version of my paper.

\end{document}